\documentclass[amsmath,amssymb,aps,prl,twocolumn,groupedaddress]{revtex4-1}
\usepackage[cp1251]{inputenc}
\usepackage{graphicx}% Include figure files
\usepackage{dcolumn}% Align table columns on decimal point
\usepackage{bm}% bold math
\usepackage{amsfonts}
\usepackage{amsmath}

\begin{document}

%\preprint{}

\title{On Oscillation Theorem for Two-Component Schr\"{o}dinger Equation}


\author{V.I. Pupyshev, E.A. Pazyuk, and A.V. Stolyarov}
%\thanks{}
\affiliation{Department of Chemistry, Moscow State University, Moscow, 119991, Russia}

\author{M. Tamanis and R. Ferber}
\affiliation{Laser Center, University of Latvia, 19 Rainis Boulevard,
 Riga LV-1586, Latvia}

%\collaboration{SHELL WE GIVE SOME e-mail here ???}

\date{\today}

\begin{abstract}
Conventional one-dimensional oscillation theorem is found to be
violated for multi-component Schr\"{o}dinger equations in a general
case while for two-component eigenstates coupled by the
sign-constant potential operator the following statements are valid:
(1) the ground state ($v = 0$) is not degenerate; and (2) the
arithmetic mean of nodes $n_1$, $n_2$ for the two-component
wavefunction never exceeds the ordering number $v$ of eigenstate:
$(n_1 + n_2)/2\leq v$.

\end{abstract}

%\keywords{CHEK THE PACS: what shell be excluded here ???}

\maketitle

% body of letter here
\renewcommand{\theequation}{A\arabic{equation}}
% redefine the command that creates the equation no.\setcounter{equation}{0}
% reset counter
% use *-form to suppress numbering
% redefine the command that creates the equation no.\setcounter{equation}{0}
% reset counter
% use *-form to suppress numbering
\setcounter{secnumdepth}{5}

\section{Introduction}\label{Sec1}

For a large number of physical applications there are useful the so called
oscillation theorems \cite{Landau, A1} that describe some remarkable properties
of solutions of the one-dimensional (1D) Schr\"{o}dinger equation. As for 1D
problems the bonded states are nondegenerate one may enumerate the wave
 functions of stationary states by their numbers $v$ in increasing energy order
($v = 0, 1, 2,\cdots$). It is known (see \cite{Landau}, Ch.XIII.3 in \cite{Reed}
 or Ch.VI, §6 in \cite{A3}) that the eigenfunction $\varphi^v$ corresponding to
the $v$-th eigenvalue $E^v$ vanishes exactly $v$ times (another
words, the $\varphi^v$ possesses exactly $v$ \emph{internal} nodes).
These statements may be clarified within the quasi-classical (WKB)
approximation (see \cite{Landau, Child}). For the many-dimensional
problems these statements are not so simple \cite{A1}.

In present study we consider the nodal properties of multi-component
wave functions for the Schr\"{o}dinger equation being interesting
within the physical problems, where adiabatic approximation shell be
used in modified forms (see e.g. \cite{Field}). For example, the
spin-orbit coupling effects also are resulted in a multi-component
diatomic vibrational Schr\"{o}dinger equation \cite{Field, Fulton}.

Let us consider the two-component Schr\"{o}dinger equation for the
$2\times 2$ matrix Hamiltonian $\textbf{H}$:

\begin{eqnarray}\label{ACC}
\mathbf{H}\vec{\varphi^v} =
\begin{bmatrix}
  h_1 & V_{12}\\
  V_{21} & h_2\\
\end{bmatrix}
\begin{pmatrix}
  \varphi_1^v \\
  \varphi_2^v \\
\end{pmatrix}
=E^v
\begin{pmatrix}
  \varphi_1^v \\
  \varphi_2^v \\
\end{pmatrix}
=E^v \vec{\varphi^v}.
\end{eqnarray}

The components $h_{1,2}$ of $\textbf{H}$ in Eq.(\ref{ACC}) have the traditional
form $h_k=-\frac{1}{2\mu} \frac{\partial^2} {\partial x^2} + U_k(x)$ with some
effective  mass $\mu$ and real-valued potentials $U_k$ (here the rotational part
of energy is included in $U_k$). The interaction operators are supposed to be
some real-valued potentials $ V_{21}(x)=V_{12}(x)$, and the energy levels $E^v$
are enumerated in increasing order, taking degeneracy into account, by
traditional vibrational quantum number $v = 0, 1, 2, \cdots$.

For molecular problems one may suppose the components
$\varphi_{1,2}^v(x)$ of a vector-function $\vec{\varphi^v}$ to be
real-valued functions defined for some region $\Omega$, being the
same for both components. It is supposed here $\Omega
=[\alpha,\beta]\subseteq \mathbb{R}^1$ for some finite (or infinite)
$\alpha$ and $\beta$. When one of the ends of the interval
$[\alpha,\beta]$, for example, $\alpha$, is finite, one supposes
that both components of solution $\vec{\varphi^v}$ obey the
Dirichlet boundary condition at this point:
$\varphi_{1,2}^v(\alpha)=0$. In any case only bound states are
considered and solution $\vec{\varphi^v}$ is supposed to be
normalized by condition

\begin{eqnarray}\label{Norm}
\langle\vec{\varphi^v}|\vec{\varphi^v}\rangle =
\langle\varphi_1^v|\varphi_1^v\rangle +
\langle\varphi_2^v|\varphi_2^v \rangle = 1.
\end{eqnarray}

The analysis of the problem in the WKB- or/and "adiabatic" approximations (of
course, being modified for the systems considered) and by the perturbation
theory (for small enough $V_{12}$) gives a wide range of possibilities. In
particular, it is easy to find some special situations when the problems under
consideration differ essentially from the traditional one-component ones. So it
seems to be interesting to describe some general statements, that are valid for
a large part of practical applications.

\section{Some examples and general notes}\label{Sec2}

In general the problems of the form (\ref{ACC},\ref{Norm}) seems to
be not essentially differs from the one-component Schr\"{o}dinger
1D-equation. It may be rewritten as a system of two equations with
the same normalization condition (\ref{Norm}):

\begin{eqnarray}\label{ACC1}
h_1 \varphi^v_1 + V_{12}\varphi^v_2 = E^v\varphi^v_1\\
V_{21}\varphi^v_1 + h_2 \varphi^v_2 = E^v\varphi^v_2
\nonumber
\end{eqnarray}

It is clear, that when interaction potential vanish ($V_{12}(x)=0$
for any $x$) the problem (\ref{ACC1}) is reduced to some
one-component eigenvalue problems for Hamiltonians $h_{1,2}$. In
this situation the Hamiltonians $h_1$ and $h_2$ with essentially
different densities of the bond states shows it clearly, that the
number of nodal points for one-component functions does not define
the number (that is, the $v$ value) of the corresponding energy
level.

If the spectra of $h_1$ and $h_2$ are identical and $V_{12} = 0$,
any energy level is degenerate one, including the ground state
level. One may also give a simple example of the situation, when
degeneracy of all the states in the two-component problem
Eq.(\ref{ACC}) appears for nontrivial $V_{12}$. Really, when $h_{1}
= h_{2}= h$ is the Harmonic Oscillator Hamiltonian on the $
\mathbb{R}^1$ axis and $V_{12}(x) = x$, it easy to transform
Eq.(\ref{ACC}) to a system of two eigenvalue problems for
1D-Hamiltonians $h + V_{12}$ and $h - V_{12}$, that describe the two
equivalent oscillators with shifted centers. Hence, all the energy
levels of the problem considered are doubly degenerate ones. Note
also, that the small perturbation of the problem above allows one to
construct the two-component problems, where the components of the
ground state have nodes.

The simple examples above described demonstrate it clearly, that the nodal
structure of the many-component wavefunctions is a much more difficult problem,
than that one for one-component 1D-system. Nevertheless, one may note, that in
all the examples presented the interaction potentials $V_{12}$ vanish, at least,
in some points. If one exclude the situations, when $V_{12}$ change its sign or
vanishes, that is, when the interaction potential $V_{12}$ \emph{is invertible},
the situation changes drastically and one may prove some interesting statements
on nodal structure of the multi-component wavefunctions.

For \emph{invertible} potential $V_{12}(x)$ one may use the system
Eq.(\ref{ACC1}) to find $\varphi_1^v$ as a solution of the 4-th order
differential equation (similar relation holds for $\varphi_2^v$):

\begin{eqnarray}\label{ACC2}
[(h_2- E^v)V_{12}^{-1}(h_1- E^v)+V_{12}]\varphi^v_1 = 0.
\end{eqnarray}

In particular, the most of statements below use the following idea, being the
result of the well known property of the Cauchy problem:

\textbf{Lemma}. For invertible potential $V_{12}(x)$ the components of any
nontrivial solution $\vec{\varphi^v}$ of the problem (\ref{ACC}) cannot be
identically zero into some open subregion of $\Omega$.

For the 1D one-component problems it is impossible that the wave
function and its gradient vanish simultaneously at some point. In
particular, near the nodal points the wavefunction changes the sign.
For the multi-component problems situation is not so simple. In
principle, for solutions of Eq.(\ref{ACC2}) it is possible that at
some point some component of the wave function only \emph{touches}
the axis. When the interaction potential does not vanish at this
point, it follows from Eq.(\ref{ACC1}) that both components
$\varphi_{1,2}^v(x)$ vanish along with their second derivatives.
This special situation is not stable with respect to small
perturbations and we ignore it here. That is, \emph{we consider as
the nodal points only the ones for which functions change their
signs}. Similar issues for one component problems for $\Omega=
\mathbb{R}^n (n>1)$ enforce to analyze the regions of a constant
sign for wave functions \cite{A1}.

\section{Useful statements}\label{Sec3}

Here we prove three useful statements that gives the background for discussions
of the nodal properties of wavefunctions in the next Section. The notations
introduced in this Section are used here and in the next Section without special
comments. In particular, we use the symbol $[n/2]$ for the integer part of the
$n/2$ value. The following relation is important for our discussions in this
text:

\begin{eqnarray}\label{nnn}
n=\left [ \frac{n}{2}\right ]+ \left [ \frac{n+1}{2}\right ].
\end{eqnarray}

An other construction is typical for the problem under consideration (see, e.g.
\cite{A1}). Let $\vec{\varphi^v}$ be an eigenvector of the two-component matrix
Hamiltonian $\mathbf{H}$ with energy $E^v$ and components $\varphi^v_1$ and
$\varphi^v_2$. Let us analyze the regions of a constant sign for the function
$\varphi^v_{1}$ (or for $\varphi^v_{2}$), that is, the intervals on
$\mathbb{R}^1$ where the function has a definite sign and \emph{changes} it at
the ends. The number $K_j$ of such regions corresponds to $n_j=K_j-1$ nodal
points of $\varphi^v_j$ ($j=1,2$).

Let us suppose that $\varphi^v_1$ has $K_1$ regions $\Omega_j^{(1)}$
$(j=1,2,\ldots,K_1)$ of a constant sign. One may define for each of regions
$\Omega_j^{(1)}$ the function $\varphi_j^{(1)}$ that equals to $\varphi^v_1$ on
$\Omega_j^{(1)}$ and equals to zero out of $\Omega_j^{(1)}$. Let us suppose that
$\varphi^v_2$ has $K_2$ regions $\Omega_k^{(2)}$ $(k = 1,2,\ldots,K_2)$ of a
constant sign and $\varphi_k^{(2)}$ are the functions similar to the ones for
the first component of $\vec{\varphi^v}$. Hence the two-component function
$\vec{\varphi^v}$ gives rise to construct the linear space $\mathcal{K}$ of
$(K_1+K_2)$ -dimensional vectors $\mathbf{C}$ defined by any set of $K_1$
coefficients $c_j$ and set of $K_2$ coefficients $d_k$ and corresponding
vector-functions $\vec{\chi}$ with components of the form

\begin{eqnarray}\label{x12}
\chi_1=\sum_{j=1}^{K_1}c_j\varphi_j^{(1)}\qquad
\chi_2=\sum_{k=1}^{K_2}d_k\varphi_k^{(2).}
\end{eqnarray}

In particular, $\vec{\chi} = \vec{\varphi^v}$ when $d_k = c_j = 1$
for any $k$ and $j$ values. Note also, that $\chi_{1,2}$ and their
gradients have to be the square integrable functions on $\Omega$ as
$\varphi^v_{1,2}$ fulfill these requirements (the sets of a zero
measure have no importance here).

Now we consider the properties of the matrix $\mathbf{M}$, defined for the above
mentioned space $\mathcal{K}$ by the vector-function $\vec{\chi}$ in the
following way:

\begin{eqnarray}\label{CMC}
(\mathbf{C},\mathbf{MC})=\sum_{j,k}(c_j-d_k)^2 \langle
\varphi_j^{(1)}|V_{12}|\varphi_k^{(2)}\rangle.
\end{eqnarray}

Our attention to the matrix $\mathbf{M}$ is explained by the
following reason. The further discussion is based on the simple
expression of energy functional $E(\vec{\chi})$ associated with the
Hamiltonian $\mathbf{H}$, for the vector-function $\vec{\chi}$ with
components of the form (\ref{x12}).

\textbf{Statement 1}. If $\vec{\varphi^v}$ is a solution of
Eq.(\ref{ACC}), then for any vector $\mathbf{C}\in\mathcal{K}$ one
may write

\begin{eqnarray}\label{st1}
E(\vec{\chi})=E^v\langle\vec{\chi}|\vec{\chi}\rangle-(\mathbf{C},\mathbf{MC})
\end{eqnarray}

\emph{Proof}. The use of a standard form for the energy functional (with kinetic
energy expressed as squared gradient of wavefunction) and integration by parts
enables to write, due to Eq.(\ref{x12}), the following relation:

\begin{eqnarray}\label{Proofst1}
E(\vec{\chi})=\sum_{j=1}^{K_1}c_j^2 \langle
\varphi_j^{(1)}|h_1|\varphi_j^{(1)}\rangle+
\sum_{k=1}^{K_2}d_k^2 \langle \varphi_k^{(2)}|h_2| \varphi_k^{(2)}\rangle\nonumber \\
+2\sum_{j,k}c_jd_k \langle
\varphi_j^{(1)}|V_{12}|\varphi_k^{(2)}\rangle,
\end{eqnarray}

where integration in each of integral in the sums is done over the regions
$\Omega_j^{(1)}$, $\Omega_k^{(2)}$ or $\Omega_j^{(1)}\bigcap\Omega_k^{(2)}$,
respectively. The use of Eq.(\ref{ACC1}) gives immediately

\begin{align}\label{Proof1st1}
E(\vec{\chi})=E^v\left[
\sum_{j=1}^{K_1}c_j^2 \langle \varphi_j^{(1)}|\varphi_j^{(1)}\rangle +
\sum_{k=1}^{K_2}d_k^2 \langle \varphi_k^{(2)}|\varphi_k^{(2)}\rangle\right]
\nonumber \\
-\sum_{j,k} c_j^2 \langle \varphi_j^{(1)}|V_{12}|\varphi_k^{(2)}\rangle
-\sum_{j,k} d_k^2 \langle \varphi_k^{(2)}|V_{21}|\varphi_j^{(1)}\rangle
 \nonumber \\
+2\sum_{j,k}c_j d_k \langle \varphi_j^{(1)}|V_{12}|\varphi_k^{(2)}\rangle
\end{align}

The use of definition Eq.(\ref{CMC}) shows it clearly, that this is
exactly expression (\ref{st1}). $\blacksquare$

It is interesting, that the number of positive/negative eigenvalues
of the matrix $\mathbf{M}$ is closely connected with numbers
$n_{1,2}$ of nodes for components of the vector-function
$\vec{\varphi^v}$. To prove it one shell note the following simple
statement.

\textbf{Statement 2}. There are no more than $(K_1+K_2-1)$ regions
$\Omega_j^{(1)}\bigcap\Omega_k^{(2)}\subset \Omega$, where the
product $\varphi^v_1\varphi^v_2$ conserves its sign.

\emph{Proof}. In general there are no more than $n_1+n_2$ internal
points of the region $\Omega$ where $\varphi^v_1\varphi^v_2$ changes
the sign. As $n_j=K_j-1$, there are no more than $(K_1-1)+(K_2-1)$
nodes for $\varphi^v_1\varphi^v_2$ , and, hence, there is not more
then $(K_1-1)+(K_2-1)+1 = (K_1+K_2-1)$ non-empty regions of the type
$\Omega_j^{(1)}\bigcap\Omega_k^{(2)}$ where the function
$\varphi^v_1\varphi^v_2$ conserves the sign. $\blacksquare$

The most important result of this Section is the following statement.

\textbf{Statement 3}. If the sign of the potential $V_{12}(x)$ is fixed, then
the matrix $\mathbf{M}$ defined by Eq.(\ref{CMC}) has \emph{no less} than
$\left[\frac{K_1+K_2+1}{2}\right]$ non-negative eigenvalues.

\emph{Proof}. According to definition (see Eq.(\ref{CMC})), the quadratic form
$\mathbf{(C,MC)}$ is determined by a sum of matrices of rank $1$, each of which
has only one nonzero eigenvalue
$2\langle\varphi_j^{(1)}|V_{12}|\varphi_k^{(2)}\rangle$,
if any. According to the Statement 2, among those eigenvalues there are no more
than $(K_1+K_2-1)$ non-zero values. A half of them is negative, while another
half is positive. In any case, according to the Eq.(\ref{nnn}) the largest
number of strictly negative integrals is no more than
$\left[\frac{K_1+K_2}{2}\right]$.

According to the minimax principle (see, e.g. Ch.VI. of \cite{A4} or
Sect.XIII.1 of \cite{Reed}), this means that no more than
$\left[\frac{K_1+K_2}{2}\right]$ eigenvalues of matrix $\mathbf{M}$
are strictly negative. Hence, the remaining eigenvalues of
$\mathbf{M}$ are non-negative and their number is no less than

\begin{eqnarray}\label{nnn1}
(K_1+K_2)-\left [\frac{K_1+K_2}{2}\right ]= \left
[\frac{K_1+K_2+1}{2}\right ]
\end{eqnarray}

as follows from Eq.(\ref{nnn}). $\blacksquare$.

\section{The Nodal Points in two-component Problem}

For 1D problems the wavefunction of the state with vibrational
number $v = 0, 1, 2, \ldots$ has exactly $v$ nodes \cite{Landau,
Reed, A3}. When $dim(\Omega )>1$, for the one-component problem one
may only prove that, for a given state, the number of regions of a
definite sign is not too large \cite{A1}. For the two-component
system considered here one may prove some analogue of these
statements. The proof is similar to the one in the one-component
case, but it requires some special modifications.

\textbf{Theorem.} For the $v$-th stationary state of the problem
Eq.(\ref{ACC}) with the interaction potential $V_{12}(x)$ of a
definite sign the number of points $n_1$ and $n_2$, where the
components $\varphi_1^v$ and $\varphi_2^v$ of the vector
eigenfunction $\vec{\varphi^v}$ change their signs, the following
inequality holds:

\begin{eqnarray}\label{nnn2}
\left [\frac{n_1+n_2+1}{2}\right ]\leq v,
\end{eqnarray}

and, hence,

\begin{eqnarray}\label{nnn3}
\frac{n_1+n_2}{2}\leq v.
\end{eqnarray}

\emph{Proof.} Let us consider the space $\mathcal{K}$ associated with regions of
constant sign for components of the wavefunction $\vec{\varphi^v}$ with energy
$E^v$, as was described in Section \ref{Sec2}. Each vector
$\mathbf{C}\in\mathcal{K}$ may define some vector-function $\vec{\chi}$ (see
Eq.(\ref{x12})). Due to Statement 1 the corresponding energy functional value
can be written as

\begin{eqnarray}\label{nnn4}
E(\vec{\chi})=E^v\langle\vec{\chi}|\vec{\chi}\rangle-(\mathbf{C},\mathbf{MC}).
\end{eqnarray}

According to the Statement 3 (remember, that $K_j = n_j+1$), the matrix
$\mathbf{M}$ has \emph{no less} than
\begin{eqnarray}\label{nnn5}
T=\left [ \frac{K_1+K_2+1}{2}\right ]=
\left[\frac{n_1+n_2+3}{2}\right
]=\nonumber\\
\left[\frac{n_1+n_2+1}{2}\right ]+1
\end{eqnarray}

\emph{non-negative} eigenvalues, each of them being associated with some vector
$\mathbf{C}$ and corresponding vector-function $\vec{\chi}$.

Let us suppose $T\geq v+2$. Then one may find \emph{at least two} independent
normalized vectors $\vec{\chi}^{(1,2)}$ (or, equivalently,
 $\mathbf{C^{(1,2)}}\in\mathcal{K}$ being orthogonal to $v$ exact solutions
$\vec{\varphi}^{0},\vec{\varphi}^{1},\ldots,\vec{\varphi}^{v-1}$
of Eq.(\ref{ACC}) with lowest energies.

For these vectors ($\mathbf{C^{(j)}},\mathbf{MC^{(j)}})\geq 0$
($j=1,2$) and, hence, $E(\vec{\chi}^{(1,2)})\leq E^v$. It follows
from the variational principle (see, e.g. Ch.VI. of \cite{A4} or
Sect.XIII.1 of \cite{Reed}) that both functions $\vec{\chi}^{(1)}$
and $\vec{\chi}^{(2)}$ are solutions of Eq.(\ref{ACC}) with energy
$E^v$ and one may construct, as a linear combination of these
vector-functions, a \emph{non-zero} solution $\vec{\chi}$ of the
Schr\"{o}dinger equation, for which, at least \emph{for one} of
regions $\Omega_j^{(1)}$ or $\Omega_k^{(2)}$, the corresponding
component of the solution ($\chi_1$ or $\chi_2$) identically equals
to zero. This is however impossible, according to the Lemma of
Section \ref{Sec1}. Hence, $T < v+2$, or, equivalently, $T \leq
v+1$. Hence, one may write
\begin{eqnarray}\label{nnn6}
\left [ \frac{n_1+n_2}{2}\right ]\leq
\left[\frac{n_1+n_2+1}{2}\right]\leq v.
\end{eqnarray}
due to Eq.(\ref{nnn5}). The usage of Eq.(\ref{nnn}) proves the
statement. $\blacksquare$

It is clear, that $(n_1+n_2)/2$ is the arithmetic mean of number of nodes for
components of the two-component solution. Hence, the Theorem statement
Eq.(\ref{nnn3}) is an analogue to the usual one-component statement for the case
of the Schr\"{o}dinger Equation in $\mathbb{R}^n (n>1)$ (see Ref. \cite{A1}).

\textbf{Corollary 1.} For the Theorem's conditions, the components of the
function $\vec{\varphi}^{v=0}$ conserve their signs into $\Omega $.

[Indeed, for $v=0$ inequality Eq.(\ref{nnn2}) means $n_1=n_2=0$].

\textbf{Corollary 2.} For the Theorem's conditions for the ground state and
$V_{12}(x)\leq 0$ the signs of $\varphi_1^v$ and $\varphi_2^v$ coincide, while
for $V_{12}(x)>0$ the signs mentioned are opposite.

\emph{Proof}. Let $\varphi_{1,2}^v$ be the components of the ground state
vector-function $\vec{\varphi^v}$. According to the Corollary 1, the signs of
$\varphi_{1,2}^v$ are fixed. Hence, for the potential $V_{12}$ of fixed sign one
may conclude that $\langle \varphi_{1}^v|V_{12}| \varphi_{2}^v\rangle \neq 0$.
The mean value of the Hamiltonian has the form

\begin{align}\label{nnn7}
\langle \vec{\varphi^v}|\mathbf{H}|\vec{\varphi^v}\rangle=\langle
\varphi_{1}^v|h_{1}| \varphi_{1}^v\rangle + \langle
\varphi_{2}^v|h_{2}| \varphi_{2}^v\rangle  \nonumber
\\+2\langle \varphi_{1}^v|V_{12}|\varphi_{2}^v\rangle.
\end{align}

If one replaces $\varphi_{2}^v$ by $-\varphi_{2}^v$ and calculates the mean
value of the Hamiltonian $\mathbf{H}$, the variational principle yields
immediately $\langle\varphi_{1}^v|V_{12}|\varphi_{2}^v\rangle < 0$. Hence, the
signs of the product $\varphi_{1}^{v}(x)\varphi_{2}^{v}(x)$ and $V_{12}(x)$ are
opposite. $\blacksquare$

\textbf{Corollary 3.} For the Theorem's conditions the ground state
is not degenerate.

[The proof is the same as in one-component case, see Ref. \cite{A1}.
Indeed, two vector-functions with components which conserve their
signs and the sign of the product $\varphi_{1}^v\varphi_{2}^v$
cannot be orthogonal. Hence, the ground state is not degenerate.]

\emph{Note}. If $dim(\Omega)>1$ one has no rights to insist on the validity of
the Theorem. However, for the interaction potential $V_{12}(x)$ of a fixed sign
one may prove Corollaries 1 - 3 by similar arguments.

\section{Conclusion}
For eigenvalue problems with matrix Hamiltonians the structure of
nodal points is much more complex than for an one-component
1D-problem, and the nodal structure of multi-component problems is
of less importance than for one-component ones. Nevertheless, for
the interaction potential of a fixed sign there are some analogues
of the oscillation theorems. In particular, for a two-component
problems for $\mathbb{R}^{1}$ axis (or a half-axis, or a finite
segment) one may prove the following statements:

(1) the ground state is not a degenerate one, and the components of
the ground-state wavefunction conserve their sign, defined by the
sign of interaction potential;

(2) for the state number $v$ (for energies in increasing order $v =
0, 1, 2, ...$) the wavefunction components $\varphi_{1}^v$ and
$\varphi_{2}^v$ have the arithmetic mean of the number of nodal
points $(n_1+n_2)/2$ not larger than $v$. (For a more precise form,
see Eq.(\ref{nnn2})).

We suppose it to be sufficient for practical applications. Though the condition
of a constant sign for $V_{12}$ seems to be too restrictive, one may note, that
in practice it is sufficient to be sure in a constant sign of $V_{12}(x)$
\emph{for the region of localization} of the wavefunction components only.

\section {Acknowledgments}

The authors are indebted to I. Klincare, O. Nikolayeva and A.
Kruzins for participation in the data processing, as well as to O.
Docenko for helpful discussions and to M. Auzinsh for useful
remarks. The support from the Latvian Science Council Grant
Nr.09.1036 is gratefully acknowledged by the Riga team.

%\begin{references}

\end{document}